\documentclass[screen, table]{acmart}
\usepackage{multirow}
\usepackage{csquotes}
\graphicspath{ {./images/} }
\AtBeginDocument{%
  \providecommand\BibTeX{{%
    \normalfont B\kern-0.5em{\scshape i\kern-0.25em b}\kern-0.8em\TeX}}}

\setcopyright{acmcopyright}
\copyrightyear{2023}
\acmYear{2023}
\acmDOI{XXXXXXX.XXXXXXX}

\begin{document}

\title{Making Data Work Count}

\author{Srravya Chandhiramowuli}
\email{srravya.c@ed.ac.uk}
\affiliation{%
  \institution{City, University of London}
  \city{London}
  \country{UK}
}

\author{Alex Taylor}
\email{alex.taylor@city.ac.uk}
\affiliation{%
  \institution{City, University of London}
  \city{London}
  \country{UK}
}

\author{Sara Heitlinger}
\email{Sara.Heitlinger@city.ac.uk}
\affiliation{%
  \institution{City, University of London}
  \city{London}
  \country{UK}
}

\author{Ding Wang}
\email{drdw@google.com}
\affiliation{%
 \institution{Google Research}
 \city{Bangalore}
 \country{India}
 }

\renewcommand{\shortauthors}{Chandhiramowuli, Taylor, Heitlinger and Wang}

\begin{abstract}
In this paper, we examine the work of data annotation. Specifically, we focus on the role of counting or quantification in organising annotation work. Based on an ethnographic study of data annotation in two outsourcing centres in India, we observe that counting practices and its associated logics are an integral part of day-to-day annotation activities. In particular, we call attention to the presumption of \emph{total countability} observed in annotation—the notion that everything, from tasks, datasets and deliverables, to workers, work time, quality and performance, can be managed by applying the logics of counting. To examine this, we draw on sociological and socio-technical scholarship on quantification and develop the lens of a ‘\emph{regime of counting}’ that makes explicit the specific counts, practices, actors and structures that underpin the pervasive counting in annotation. We find that within the AI supply chain and data work, counting regimes aid the assertion of authority by the AI clients (also called requesters) over annotation processes, constituting them as reductive, standardised, and homogenous. We illustrate how this has implications for i) how annotation work and workers get valued, ii) the role human discretion plays in annotation, and iii) broader efforts to introduce accountable and more just practices in AI. Through these implications, we illustrate the limits of operating within the logic of total countability. Instead, we argue for a view of counting as partial—located in distinct geographies, shaped by specific interests and accountable in only limited ways. This, we propose, sets the stage for a fundamentally different orientation to counting and what counts in data annotation. 
\end{abstract}



\maketitle

\section{Introduction}

As AI’s potential and scope reach and cross new horizons, data annotation continues to be an indispensable and crucial part of its development \cite{sambasivan2021everyone, hutchinson2021towards, paullada2021data}. Data annotation (or data labelling, often used interchangeably) refers to a range of activities such as data cleaning, tagging, classification, comparison, and verification required to build training datasets for AI/Ml models to be trained on. The ‘intelligence’ that AI-based systems typically provide is developed by processing and ‘learning’ from the patterns and labels in large-scale training datasets. These large-scale datasets required to develop AI technologies are rarely available readily. Data has to be collected or generated, cleaned, curated, classified, annotated, and verified before AI-based systems can harness them for “intelligent” output. The significant effort involved in data cleaning, annotation, and verification is largely carried out by human data workers \cite{wang2022whose, roberts2021your, gray2019ghost}.

In a rapidly growing global industry of data work \cite{gray2019ghost, irani2015cultural, sambasivan2021everyone, martin2014being}, thousands of annotation workers create training datasets, validate model outcomes and mimic computational responses \cite{tubaro2020trainer} through digital labour platforms \cite{posada2022coloniality, poell2019platformisation} and outsourcing companies \cite{ahmad2022moderating, miceli2020between, wang2022whose} to sustain AI’s research, development and use. Tasks such as image or text labelling or content moderation are distributed to actors and agencies, removed or distanced from those geographic regions so often celebrated for the innovation in AI. This work is predominantly carried out in the “majority world” \cite{amrute2022primer}, where it is increasingly framed as an ‘opportunity’ for workers, families, and communities experiencing economic crises, political unrest, or social stigma \cite{graham2017digital, posada2022embedded}. Against this backdrop, there has been a growing interest within CSCW/HCI and adjacent communities to better understand the processes of data annotation and their implications for AI, particularly, what and who is involved in this work. Here, critical scholarship in this regard \cite{irani2015cultural, bilic2016search, roberts2019behind, gray2019ghost, posada2022embedded, miceli2020between, wang2022whose, chandhiramowuli2023match} has revealed the difficult working conditions, unfair employment relations, and power asymmetries within data annotation work that fuel the 21st century’s algorithmic systems.  

We build on this body of work to present an ethnographic study of data annotation at two outsourcing annotation centres in India. In this paper, we specifically focus on the pervasive logics of counting observed in the everyday work practices of data annotation. In particular, we call attention to the normalised assumption of \emph{total countability} observed in annotation – the notion that everything, from tasks, datasets, and deliverables, to workers, work time, quality, and performance, can be managed by applying the logics of counting. We use the term counting to refer to techniques of enumeration and quantification that are used to render events, processes, and contexts in routine and standardised ways. This understanding builds on a line of scholarship in sociology, science and technology studies, and CSCW that investigates and engages with counting practices in varied settings \cite{latour1987science, rose1991governing, miller1987accounting, power2004counting, verran2001science, bowker2000sorting, bowker2001pure}. 

We use the theorising of counting to examine data annotation as not merely a series of methodical tasks performed by individuals but a structural activity that constitutes the wider practices and the logics that undergird them. In this way, counting is seen as a form of \emph{structuring} \cite{orlikowski2002s, 10.1145/1958824.1958861}, the means through which people, organisations, and ways of knowing are structured through routine practice. The theorising on quantification as structuring helps us develop an analytical frame to consider the actors involved, the practices undertaken, the mechanisms or technologies used in counting, and the ideas or logics they underpin within annotation. Bringing this orientation to bear in our analysis, we see that the pervasive logic of counting observed in annotation serves to situate control and authority amongst a few actors, echoing the findings in \cite{10.1145/3555561} and \cite{wang2022whose}.

To extend this theorising and pay particular attention to the ways power and authority accompany structuring, we think with the term \emph{“regime of countability”} used by Geoff Bowker and Susan Leigh Star \cite{bowker2001pure}. For Bowker and Star, the term helps foreground the reliance on counting and what they call the “crisis of quantification,” particularly, in the ongoing political and democratic instability in the US. They argue that counting underpins citizens' existence—that “a modern state needs to conjure its citizens into such a form that they can be enumerated” (p. 423). To be a good citizen, then, one has to neatly fit into a countable classification (e.g., age, race, gender, etc.); “who can’t be counted are the people who don’t count.” (p. 423). Such counting, though, is totalising to the extent that everything is presumed countable, and the structures and technology used to produce the counts are no longer visible or accountable. We apply this idea of a \emph{regime of counting} applied in modern governmentality to data annotation, where the prevailing logic of counting rules over what and how annotation works come into being. 

Informed by our orientation to counting as a structural activity as well as by our observations of its dominance in annotation, the overarching question that motivates this research is: What specific work does counting do in data annotation for AI? In operationalising this research question, we are guided by the following additional questions: i) what is the role of counting in data annotation? ii) what actors, practices, and technologies are involved in operationalising counting within annotation work? iii) what specific counts are produced and normalised (and what alternatives are marginalised or erased) as a result of the interests, priorities, and concerns surrounding data annotation?

We address the research questions by drawing on an ethnographic study of data annotation that we conducted in India between June - August 2022. We visited two annotation centres (or offices) in India that perform outsourced data work for AI clients. The centres were located in smaller cities and towns, providing employment to young graduates from the local area\footnote{This form of employment generation in smaller cities, towns and rural areas through outsourced tech work is referred to as impact sourcing.}. Both centres offered a range of data services, including building training datasets, verifying model outcomes, and real-time support to correct any erroneous outputs of AI systems. Through these services, we found that annotation workers undertake data tasks whose verdicts serve as training data for AI models and systems, continually fed back to improve model performance and inform system enhancements. For this reason, we use the term data annotation throughout the paper to refer to the range of data services offered to AI clients. 

Through our empirical study of data annotation, we show the central role of counting. In the day-to-day activities of the annotation centres, counting forms an integral part of organising annotation work and managing the workers. However, we find that counting practices are far from passive or inert—mere representations of labour or tools to enumerate productivity. Clients wield a disproportionate influence over the processes and outcomes of annotation through counting. Consequently, they control what gets counted and accounted for. The fallout of their priorities and interests can then be traced through the tensions and erasures in counting. We illustrate how this affects i) the value and valuing of annotation work, ii) the recognition of the human discretion necessitated in annotation, and iii) broader efforts to introduce accountable and more just AI systems. Through these implications, we illustrate the limits and dangers of operating within the logic of \emph{total countability}. Instead, we argue for a view of counting as partial—shaped by specific interests and incapable of accounting everything. In recognising counting as partial and not totalising, we see possibilities for better counting, not only quantitatively but also in ways that reconsider what counts.

Through this study, we contribute to two areas of CSCW research: empirical studies of data annotation and theorising of counting. First, our work offers an in-depth engagement with an important theme for consideration in the study of data annotation, that is, counting. Our study offers an up close view of how data annotation is constituted through counting practices and reflects on its implications for AI system building. As we witness a surging interest in the adoption of AI, annotation — a foundational aspect of AI production — warrants close scrutiny, particularly how it is structured and conducted. Our contribution here is to unpack the structuring work afforded and enacted through the regime of counting in data annotation. We aim to recover annotation from the reductive framing and seek alternative approaches to knowing and doing annotation. We aim to show that counting, as a regime, is not neutral but, in Helen Verran’s terms \cite{verran2001science}, \emph{world building}. Instead of valuing people’s unique capacities for judgement, discretion, and collaboration, the counting regime limits the registers of action and interaction for annotators, for the data and AI sector, and ultimately for AI systems. Perversely, the logic of counting in data work creates the conditions for humans to behave more like machines and offers little opportunity for machines to learn from people. This, we argue, limits the possibilities; the chances of other forms of knowing and being are reduced, relegated, and pushed to the margins. In this, our work engages closely with the themes of power and authority developed in earlier CSCW work \cite{miceli2020between, 10.1145/3555561, sambasivan2021everyone, wang2022whose} and extends these themes by contributing the specific case of counting. Further, our study adds an empirically rich field study of data annotation work, particularly from the Indian outsourcing context, which has not yet been examined closely. Our work complements and adds to the growing body of research on data annotation as conducted in other parts of the world. Finally, we contribute to the theorising on quantification by developing the analytical lens of \emph{‘regimes of counting’}; this builds on sociological and socio-technical scholarship on quantification and brings them together in a novel conception that confronts and grapples with the idea of ‘countability.’ This holds relevance to research on data work but also to other areas witnessing rapid adoption of metrics, datafication, and algorithmic management.

\section{Related Work}
In this study examining counting practices in operation in data annotation, we begin by discussing the body of work emerging around the labours of data annotation for AI. We then build on a line of scholarship in sociology and science and technology studies (STS) which investigates and problematises counting and quantification. 

\subsection{Data Annotation for AI}

Machine learning, neural networks, computer vision, natural language processing and robotic process automation—all broadly grouped under the label of AI—are increasingly being deployed in a wide range of contexts from agriculture and mining to healthcare and education. AI’s ability to forecast, personalise, and automate is leveraged both to address legacy challenges in these sectors and explore new possibilities. The ‘intelligence’ that AI-based systems provide is built by processing large scale datasets. These datasets require extensive preparation, cleaning and annotation—processes that are an indispensable part of AI development and largely carried out by human workers \cite{sambasivan2021everyone, paullada2021data, hutchinson2021towards}. Large corporations and start-ups alike crowdsource or outsource data tasks such as image labelling, segmentation, object recognition, and language annotation to train and enable “automated” systems \cite{gray2019ghost, irani2013turkopticon}. Data annotators not only provide clean, labelled data, they also validate the model output and stand-in for the AI system at points of breakdown or failure.  

Despite data annotation being integral to AI systems, there is a model-centric view of AI that celebrates and accords high status to algorithmic aspects of AI \cite{sambasivan2021everyone}. Sambasivan \cite{sambasivan2022all} has argued that this overt emphasis on models overshadows the contributions made by people and more worryingly, erases those drafted into providing data support for these systems. This overshadowing is evident in the claims of model neutrality that shift the responsibility of bias in AI systems on to the ‘subjectivities’ of data workers who label the data. However, Miceli et al \cite{miceli2022studying} and Wang et al \cite{wang2022whose} challenge these claims by revealing the power asymmetries in data annotation that prevent data workers from providing any meaningful input or introducing their own subjectivities. Referring to a data production dispositif, Miceli and Posada \cite{10.1145/3555561} demonstrate how the work practices, tools, platforms and task instruction artefacts all come together to configure data work in response to the “voracious” demand for more, cheaper, and specialised datasets to support a rapidly growing AI industry. Indeed, this is bolstered by critical scholarship from Bilić \cite{bilic2016search}, Gray and Suri \cite{gray2019ghost}, Irani \cite{irani2015cultural}, and Roberts \cite{roberts2019behind}. These works powerfully reveal the difficult working conditions, unfair employment relations, and power asymmetries within data annotation work that fuels the 21st century’s algorithmic systems. By design, crowdsourcing platforms, which are popular sites for annotation work, restrict the bargaining power of data workers, often resulting in a ‘race to the bottom’ in wage rates and working conditions \cite{graham2017digital}. 

In recent years, there has been a shift from crowdsourcing annotation tasks to outsourcing them to third-party vendors ostensibly for better quality control. In this arrangement too, annotation tasks are heavily shaped by the interests, values, and priorities of those who design, request and pay for the work, with very limited agency given to those actually doing it \cite{miceli2020between, wang2022whose, chandhiramowuli2023match, le2023problem}. A growing workforce, largely distributed across the global south, forms a crucial link in the feedback loop within AI/ML development, and yet this work is done amidst concerns of limited autonomy, fears of redundancy, and deskilling. In this paper, we examine outsourced data annotation work conducted in India, paying particular attention to the work practices and organisational settings within which annotation is conducted\cite{graham2017digital}. Our focus on counting seeks to extend the existing scholarship, drawing attention to a particular form of structuring that enacts the conditions and practices of data work.

\subsection{Quantification and Ways of Knowing}

There is a rich body of work in CSCW, HCI and adjacent areas of socio-technical research  on the adoption of counting practices in the workplace. At the turn of the twentieth century, the practice of standard costing and budgeting played a crucial role in the construction of the individual worker as a manageable and efficient entity. Taken together with the principles of scientific management, standard costing contributed to notions of efficiency of individual workers being expressed in monetary terms \cite{miller1987accounting}. More recently, over the last two decades, there has been a rising adoption of data related technologies towards the quantitation of work in various domains \cite{moore2016quantified}. The use of audience metrics in journalism quantifies audience engagement to inform editorial decisions and journalistic practice \cite{christin2018counting}. Data-intensive sports analytics produce metrics and data points that shape recruitment of players, organisation of teams and game strategies \cite{beer2015productive}. Customer ratings and review systems render platform worker performance in numbers and form a part of a broader system of algorithmic management \cite{wood2019good}. Wearable self-tracking technologies generate large amounts of data on workers to deduce patterns of productivity, well-being and job satisfaction \cite{moore2016quantified}. In these cases and many others, counting renders rich social contexts as standardised traces by drawing on and feeding into the normative scheme of numerical objectivity. 

The practices of counting and quantifying to make sense of an unknown context, population or place have a long and complex history, entangled in colonial power and exertion of control. William the Conqueror launched the Great Survey in eleventh century England to record the value each settlement brought to its lord, counting the tenants and resources (including land, livestock and manpower) from which the value was derived \cite{godfrey1996accountability,oldroyd1997accounting}. This data was compiled into what is famously called the Domesday Book, documenting the efforts and outcomes of counting to consolidate the regime’s power and assert its fiscal rights \cite{mcdonald2005using, mennicken2019s}. Counting practices and statistical knowledge were also instrumental in producing and reifying colonial understandings of caste in South Asia \cite{cohn1984census, cohn2020colonialism, samarendra2011census, kalpagam2000colonial} and “tribe” in Nigeria \cite{van2004establishing}. The enumerative practices in these contexts aided the assertion of authority over events and processes distant from the centres of power by inscribing them in standardised forms that could be transported back to be “aggregated, compared, compiled and calculated about” \cite{rose1991governing}. In doing this work of translating and transporting a qualitative world into information, data and statistics served as an apparatus of domination \cite{hacking1990taming}.

Contemporary counting practices continue to be caught in and shaped by politics of domination and authority \cite{davis2012governance, merry2016seductions, rottenburg2015world}. Yet, quantification remains seen as an objective approach to governance; a neutral, impartial exercise that can reveal the truth. In his long standing sociology of numbers, Porter \cite{porter1996trust} explains that establishing a domain of objectivity became particularly relevant in the face of social transformations that increased population mobility and expanded trading into new markets. In the absence of old bonds and existing trust, quantification as the new substitute allowed decision making to be anchored in rule following. Measurements are replicable, not dependent on when, where and by whom the measurement is done \cite{power2004counting}. Thus, numerical rules (of counting) privilege impersonality over wisdom or experience in decision making. 

By standardising events and processes to render them enumerable, counting fails to account for the complexities of that which is being counted. This inability to accommodate contextual complexities is significant for two reasons. The first is an epistemological concern that counting produces a “realisation” of theoretical categories in the said context rather than the “representation” of the context \cite{latour1987science, hacking1983representing}. The second reason concerns the implications of this reduction. Sociologist Starr argues that the reduction of complexity, though inherent to counting, cannot be viewed as ideologically or theoretically innocent \cite{starr1987sociology}. The framing of the numerical enquiry, systems of classification, methods of measurement, frequency of measurement - all of these choices embody and embed the expectations, beliefs and concerns of those responsible for the counting \cite{rose1991governing}. Quantification’s ultimate reduction of complexity — a single quotable, comparable and calculable number to represent a complex phenomenon — poses the risk of normalising the single figure as that which is being counted.    

The above strands of theorising on quantification engage with themes like domination, control, objectivity and complexity that we find relevant to our study of counting practices in data annotation and help us develop an analytical frame to examine the work we report on here. Specifically, they inform how we consider the role of counting as a structural activity, bringing into focus the actors involved, the practices undertaken, the mechanisms or technologies used, and the ideas or logics they underpin. We bring these different but related themes of analysis together through the frame of \emph{‘regimes of counting’}—a term used by Bowker and Star \cite{bowker2001pure} to describe the turn towards ubiquitous counting and unlimited countability in American politics and governance—the belief that everything, be it citizens, votes or diseases, can be measured and accounted for through counting, classification and statistical methods. What is key here in their articulation, and important for our analysis, is the totalising nature of this turn, where everything is presumed to be countable. Bowker and Star argue that by blurring the work and technologies involved in counting into the background, it becomes possible to entertain the imaginary of \emph{total countability}. This then fosters a dependence on counting to navigate complex social contexts (like elections, census, healthcare, education, immigration) and account for them in decision-making processes. Underlying this reliance is a perverse logic at play: that what is countable is what matters and consequently, the uncountable, does not count. In making this explicit, Bowker and Star’s commentary is an invitation to challenge both the practices of pervasive counting as well as visions of total countability. 

Rather than being a concretely defined term, \emph{‘regimes of counting’} allows us to develop an analytical frame to grapple with the reliance on counting and countability. It lets us bring attention to the specific actors involved in counting, work practices of producing counts and the technologies involved. It opens up the opportunity to consider how questions of authority, dominance, objectivity and complexity come to bear within counting regimes. Helping to refine this analytical frame, we also put this line of thinking in conversation with Helen Verran’s work in \emph{Science and An African Logic} \cite{verran2001science}. Verran’s thought-provoking and careful examination of the entangled relations between colonial geographies, language and experimental apparatus show counting to be foundational to not only ways of knowing but to ways of being in the world. Alongside the idea of regimes of counting, this ontological inquiry invites us to be attentive to the particular ways in which counting regimes might conjure the worlds they enumerate, and in doing so normalise very narrow ideas of what matters or ‘what counts’. 

It is this critical orientation to counting that we use to develop a focused reading of data work and to invite the questions: i) what is the role of counting in data annotation? ii) what actors, practices and technologies are involved in operationalising counting within annotation work? iii) what specific counts are produced and normalised (and what alternatives are marginalised or erased) as a result of the interests, priorities and concerns surrounding data annotation? 

In examining these questions, we present an in-depth engagement with the prevailing practices of counting in data annotation, and reflect on its implications for data work and the AI systems they feed into.

\section{Methods}

\subsection{Fieldsite}

We adopted an ethnographic approach to study the work practices of data annotators. We conducted the study at two annotation centres affiliated with Data Futures\footnote{Names and identifiable details of people, organisations and projects have been pseudonymised.}, a company that provided third-party data annotation services to support AI/ML products and services. While headquartered in what is considered India’s technology capital, Bangalore, Data Futures is affiliated with 6 annotation centres, operating in smaller cities and towns in India. At the time of writing, Data Futures HQ was responsible for procuring data annotation projects and outsourcing them to the annotation centres. Their clients varied, ranging from start-ups to large corporations building AI products and services, based in India, UK, USA as well as other parts of the world, and operating in domains such as ecommerce, security, navigation, and retail shopping. The study was initiated through a partnership between the researchers and Data Futures. Data Futures was chosen because it offered an opportunity to understand the practice of data annotation in action, aligning with our research interests. Data Futures due to its distributed model and ambition to grow wanted to understand the best practices happening in data centres that made their organisational practices. We shared our findings with Data Futures after analysis.

The two Data Futures centres where we conducted our fieldwork are referred to here as Veera DF and Chandra DF. In both centres, there were multiple annotation projects that were ongoing during our visits. These included a project to verify automated recognition of road signs and turn restrictions for a mapping system; a product matching project to compare and match products to aid price comparisons; a mapping project to draw polygons over buildings and label their attributes; and a live support project to support automated, till-less shopping. Some of these were large scale projects involving hundreds of annotators and over fifty thousand annotations per day; some were ‘PoC’ (proof of concept) projects to build training datasets for nascent AI models; some involved validating model outcomes, while others required annotators to stand-in for AI\footnote{For instance, in the automated, till-less shopping project, this included annotators completing tasks such as identifying shoppers that the AI failed to identify or billing shopping carts that weren’t accurately billed by AI. We draw on more vignettes and observations from this project in the later sections.}, completing the job where the technology couldn’t. For the purposes of this paper, we use the term “annotation” to refer to this broad range of data services offered to clients building AI/ML systems. 

Each annotation project consisted of a team of entry level annotators, referred to as agents or analysts, and senior annotators who took the role of quality check (QC) and quality assurance (QA) analysts. The prospect of engaging in office work in high-tech domains attracted young, enthusiastic and educated graduates in the local area. The entry level annotators were typically in their 20s, having recently completed bachelors or masters degrees, and coming from the towns and surrounding areas where the annotation centres were located. At both centres, more than 50\% of employees were female. Depending on the size of the project, there could be multiple teams working on the same project, each headed by a team lead (TL) and supported by one or two assistant team leads (ATLs). The TLs reported to an assistant manager who handled multiple projects. TLs were primarily responsible for the day to day activities while managers liaised with the clients on a regular basis with project updates, clarifications and other reports. In many cases, TLs and managers were also senior employees at the centre who had joined as annotators and moved through the ranks of the organisation. We visualise the structure of a typical team and hierarchy of roles in Fig.1. 

\includegraphics[scale=0.35]{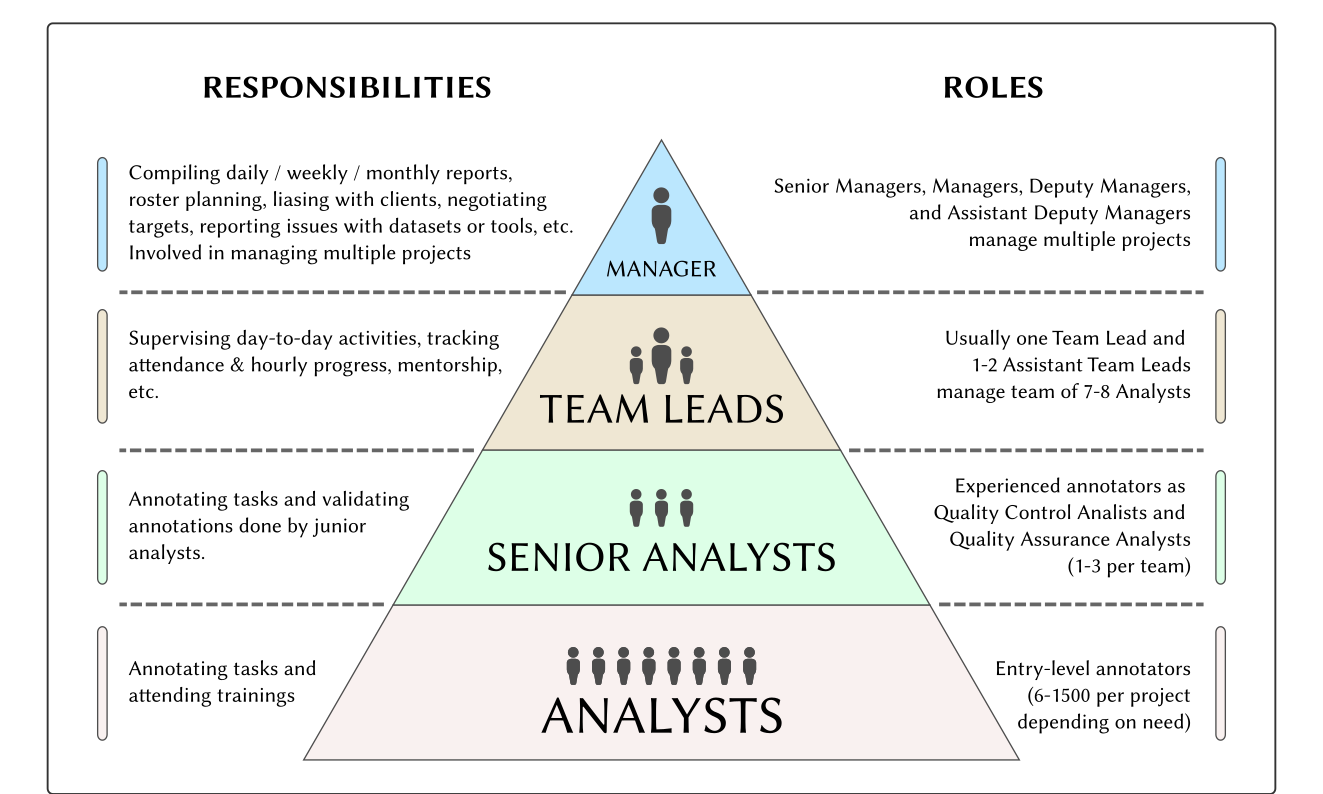}
\begin{figure}
\centering
\small
\caption{Hierarchy of roles and corresponding responsibilities of actors in an annotation centre}
\medskip
\end{figure}

\subsection{Data Collection}

Our study began with identifying annotation projects that we could observe and follow at the two Data Futures annotation centres. With support and input from the senior managers working at the centres, we identified five projects to concentrate on. These projects vary in the annotation tasks, tools used, domains, clients requesting the work and the scale of operation. PoC projects typically tend to be small-scale, involving a 5-6 member team. In larger projects, the number of annotators could be as high as 1500 with tens of teams, each consisting of about 8-10 members.

The fieldwork involved observations and interactions—occurring between June to August 2022 over four week-long visits made by the first author to the two data centres, amounting to 160 hours of observations. During this time, the observations and interactions spanned across 8 annotation teams. This time included following the annotation work by sitting alongside and observing annotators as they performed their data tasks. It also included attending training sessions for newly recruited annotators, and participating in several team meetings, task walk-throughs and regular debriefs communicating updates from the clients on new instructions or changes in priority. The observations were supplemented by interactions with 30 annotators, 8 team leads and 7 managers—taking place on the office floor and at participants’ workstations while they worked. The interactions occurred over impromptu conversations during work, in coffee breaks, lunch table chats and bus rides to and from the annotation centre sites. 

Through these observations and interactions, we gained detailed insights into the real world contexts in which annotation work occurs, the nature of data tasks, the varied experiences with annotation tools and the structures that governed this use, and the challenges, breakdowns and workarounds that shaped the everyday experiences of doing annotation. To understand the organisational structures and practices that influenced annotation, including hiring practices, negotiations with clients and onboarding new projects, we conducted 9 in-depth interviews with senior team leads and managers at both the centres who had been with the organisation for at least five years. These interviews were semi-structured, recorded with informed consent, and lasted 70 minutes on average. 

All interactions and conversations including interviews were in a mix of English and the local languages spoken at the centres. Access was gained to the two centres by seeking the permission\footnote{We signed a legally binding agreement with the headquarters, outlining our access to the centres for research purposes and requirements for non-disclosure and confidentiality of all sensitive, business-related data. Besides this, the study was also approved by our institutional review board.} of the headquarters, Data Futures, and subsequent coordination with the centre representatives to conduct the field study. Further, explicit permission was sought from annotation workers as we came in contact with them during data collection  through interactions, observations and interviews. 

\subsection{Data Analysis}

The data collection described above resulted in copious handwritten field notes, recorded interviews, and pictures and illustrations from the field sites. The interviews were transcribed and the field notes were written up in English, translated from local languages when required. All sensitive and identifiable details, including names of persons, projects, clients, and tools were pseudonymised. The anonymised, translated field data was shared among the authors. 

Our analysis draws inspiration from a long tradition of inductive approaches and ethnographic research in CSCW \cite{randall2007fieldwork, randall2021ethnography, wang2020please, karusala2020making, muralidhar2022between, taylor2003gift}. This tradition of inquiry focuses on the ways in which actors order their activities and how they draw on artefacts, technologies and their knowledge, skills and expertise to conduct such ‘work’. The emphasis here is less on how often something is said by participants or an activity occurs, and more on how the actions are made sense of, collectively worked on and come to shape in-situ, social practices. Lucy Suchman’s canonical \emph{Plans and Situated Actions} \cite{suchman1987plans} is perhaps the most well-known example of such empirical work in CSCW and HCI, but the orientation has emerged and indeed evolved to become a staple in qualitative research in both areas. 

In practice, our analysis involved repeated listening and readings of transcripts and fieldnotes to identify and pay close attention to where actors made their thinking available through both their actions and spoken interactions. Analysis was conducted individually and together by the paper’s authors. The aim was to understand the situated nature of data practices and how participants contributed to and made sense of ongoing work. Immersion in the data—through reading, writing, and discussion—helped uncover gaps in understanding or new salient phenomena that, in turn, informed subsequent periods of data collection and resulted in the emergence of the themes around annotator work and counting that form our findings. For instance, initial visits to the two data centres revealed the centrality of monitoring and tracking across projects. As we repeatedly revisited the field notes and transcriptions across both sites, we paid attention to the seemingly ordinary details of related activities such as marking attendance or recording hourly task counts. Altogether, the individual and collective iteration and  progression of fieldwork helped to compose what Paul Dourish \cite{dourish2001action} has characterised as \emph{thick description}—“to describe not just what the members of that culture do but what they experience in doing it; why it is done and how it fits into the fabric of their daily lives” (p. 59).

It was through this process that our attention was drawn to the many numbers and counts that defined annotation work practices. We probed this theme through phases of thinking and writing informed by scholars such as Helen Verran \cite{verran2001science}, Geoffery Bowker and Susan Leigh Star \cite{bowker2000sorting,bowker2001pure}, and Theodore Porter \cite{porter1996trust}, whose works bear a strong influence on CSCW and HCI. This allowed us to deepen and refine our insights into numbers and counting not just in practical terms but also their structural qualities—that is, in terms of how they structure not only what is done, but how to think through, organise and judge the varied practices surrounding annotation.

\subsection{Positionality}

All authors are HCI researchers, affiliated with institutions in the global north. The fieldwork was conducted by the first author, who was located in India during the study. The first author was born and raised in the broader region in India where the study was conducted, and is also fluent in the languages spoken in the two centres. This helped situate the observations and practices within a broader cultural context. This author’s past experience with doing annotation work (through participant observation in earlier research) proved valuable in both paying attention to nuanced aspects of annotation work but also building rapport with the annotators. Within the female-majority workplaces, the first author’s identity as a woman allowed her to make individual connections with annotators. 

Despite sharing these similarities in identity and experiences, we recognise that our work cannot be distanced from the privileges afforded by us as researchers affiliated with global north institutions examining annotation contexts in India. This reflexivity informed our efforts during fieldwork to be open and transparent about our research aims, interests and outcomes, including its limits. This also shaped our approach to the analysis and writing up of the study, where we have taken intentional efforts to exercise care while examining our participants’ work lives and presenting their lived experiences.

\section{Findings}

At both the annotation centres, each annotation project differed in the types of tasks carried out, the kinds of AI systems they fed into, contexts or domains to which they were applied, and annotation tools they used. Across these varied annotation projects, clients, tools and contexts, one practice found common relevance and was intricately woven into the annotation work itself — counting. In the sections that follow, we elaborate on the forms of counting that permeate annotation work in the field sites. We show how counting is an essential part of doing data annotation, from roster planning to reporting each day’s work. It permeates the day to day annotation activities, as the de facto language in which annotation is not only described but also conceptualised. However, the many counts that make up annotation’s vocabulary are not a given; they are produced through negotiations that influence other kinds of counts. In the final section of the findings, we show where the limits of such counting arise. 

\subsection{Counting as necessary practice}

We begin by showing how counting, in its most basic form, is integral to the management of data annotation practices. We present observations and vignettes on two day-to-day activities that were instrumental to organising and conducting annotation work — roster planning and end-of-day reporting. Through these activities, we highlight the significance of counting in organising annotation work within the outsourcing supply chains. Here, we find counting practices to be taken for granted, whether it is to quantify the labour required to meet project demands or enumerate progress to fulfil clients’ contracts.

At Chandra DF, a team of nearly 100 annotators provided live support for a client that developed AI systems for automated, cashierless checkout in retail stores\footnote{The stores were predominantly in the global north.}. The system allows shoppers to pick items for purchase and exit the store, without having to queue at checkout tills. Customers are instead automatically billed, by the AI-based system that has tracked and identified the items they have picked and left a shop with. The live support team at Chandra DF was responsible for near real time assistance, stepping in virtually to verify shopping carts and ensure accurate billing, when the AI failed to do so. The near-real time tasks carried out by this team ensured smooth operations of AI-based checkout, while also feeding into its training datasets. This support project was headcount-based\footnote{Annotation projects typically followed a transaction volume-based model, in which the client paid for a fixed volume of work to be delivered (say 50,000 tasks each day) or a headcount-based model, where they paid to contract a fixed number of people (say 75 annotators) to work full time on their annotation tasks.} — the client contracted a fixed number of people to work on the project, across three shifts (morning, mid-day and night) covering 24 hours a day and all 7 days a week. Here counting was necessary and integral to how the work was organised. 

Every week, the client provided the staff count required in each shift. If the project headcount was 100, the client dictated how the 100 annotators were split across 3 shifts each day of the week. And this changed every week, based on the anticipated footfall at the stores. For instance, they required the highest count for the night shift (10PM - 7AM IST) since this period overlapped with busy shopping hours in North American stores. Fridays and Saturdays were busier than, say, Tuesdays and Wednesdays (as illustrated in Table 1). Moreover, during sporting events on weekends, they needed more annotators logging in for live support from India to ensure hassle-free shopping for fans rushing in and out of the checkout-free stores in soccer stadiums.

\begin{table}
\begin{tabular}{@{}|l|c|c|c|c|c|c|c|@{}}
\toprule
\textbf{Shift} &
  \multicolumn{1}{l|}{\textbf{Mon}} &
  \multicolumn{1}{l|}{\textbf{Tues}} &
  \multicolumn{1}{l|}{\textbf{Wed}} &
  \multicolumn{1}{l|}{\textbf{Thurs}} &
  \multicolumn{1}{l|}{\textbf{Fri}} &
  \multicolumn{1}{l|}{\textbf{Sat}} &
  \multicolumn{1}{l|}{\textbf{Sun}} \\ \midrule
\textbf{Morning}   & 31 & 35 & 35 & 34 & 32 & 33 & 32 \\ \midrule
\textbf{Mid}       & 17 & 17 & 18 & 16 & 16 & 17 & 15 \\ \midrule
\textbf{Night}     & 37 & 35 & 33 & 34 & 42 & 39 & 43 \\ \midrule
\textbf{Off-count} & 11 & 9  & 10 & 12 & 6  & 7  & 6  \\ \midrule
\textbf{Total}     & 96 & 96 & 96 & 96 & 96 & 96 & 96 \\ \bottomrule
\end{tabular}
\caption{An illustration of the headcount planning sheet. The table shows the shift-wise headcount that the client required each day across the three shifts. Project managers used this as the basis to plan the week’s roster, computing the work days and rotational offs for each annotator in each shift. The morning shift was 7AM - 4PM, mid-day shift 1PM - 10PM and night shift 10PM - 7AM. The off-count row seen in the table refers to the number of people who could be given their weekly time-off; this is factored in by the project’s manager based on the client’s shift-wise requirements. }
\label{tab:my-table1}
\end{table}

Once the client shared the required headcount per shift, the project manager and shift TLs would draw up a roster plan for who would work on which days. Each annotator received 6 days off in a month, given as one and two days off per week alternatingly. Overlaying this leave count onto the headcount, the roster plan for each shift dictated the work days and the off days for each annotator (see Table 2). This roster was not static; there could be unplanned leave due to sickness or emergencies, requiring the TL to work in a buffer count, a few extra people in each shift. Every week the TL updated this roster based on the changing headcount requirements of the client, while also weaving in the leave schedule of annotators. The annotators could swap their day off with a colleague if needed and with prior approval of the manager, further complicating and necessitating the meticulous counting and tallying. Every shift began with the TL marking attendance, ensuring that the required headcount was met.

\begin{table}
\begin{tabular}{@{}|l|l|lllllll|lllllll|@{}}
\toprule
\multicolumn{1}{|c|}{} &
  \multicolumn{1}{c|}{} &
  \multicolumn{7}{l|}{\textbf{Week of July 11 - 17, 2022}} &
  \multicolumn{7}{l|}{\textbf{Week of July 18 - 24, 2022}} \\ \cmidrule(l){3-16} 
\multicolumn{1}{|c|}{\multirow{-2}{*}{\textbf{Name}}} &
  \multicolumn{1}{c|}{\multirow{-2}{*}{\textbf{EmpID}}} &
  \multicolumn{1}{l|}{\begin{tabular}[c]{@{}l@{}}11\\ M\end{tabular}} &
  \multicolumn{1}{l|}{\begin{tabular}[c]{@{}l@{}}12\\ T\end{tabular}} &
  \multicolumn{1}{l|}{\begin{tabular}[c]{@{}l@{}}13\\ W\end{tabular}} &
  \multicolumn{1}{l|}{\begin{tabular}[c]{@{}l@{}}14\\ Th\end{tabular}} &
  \multicolumn{1}{l|}{\begin{tabular}[c]{@{}l@{}}15\\ F\end{tabular}} &
  \multicolumn{1}{l|}{\begin{tabular}[c]{@{}l@{}}16\\ Sa\end{tabular}} &
  \begin{tabular}[c]{@{}l@{}}17\\ Su\end{tabular} &
  \multicolumn{1}{l|}{\begin{tabular}[c]{@{}l@{}}18\\ M\end{tabular}} &
  \multicolumn{1}{l|}{\begin{tabular}[c]{@{}l@{}}19\\ T\end{tabular}} &
  \multicolumn{1}{l|}{\begin{tabular}[c]{@{}l@{}}20\\ W\end{tabular}} &
  \multicolumn{1}{l|}{\begin{tabular}[c]{@{}l@{}}21\\ Th\end{tabular}} &
  \multicolumn{1}{l|}{\begin{tabular}[c]{@{}l@{}}22\\ F\end{tabular}} &
  \multicolumn{1}{l|}{\begin{tabular}[c]{@{}l@{}}23\\ Sa\end{tabular}} &
  \begin{tabular}[c]{@{}l@{}}24\\ Su\end{tabular} \\ \midrule
\textbf{Devi} &
  2034 &
  \multicolumn{1}{l|}{\cellcolor[HTML]{00FF00}P} &
  \multicolumn{1}{l|}{\cellcolor[HTML]{00FF00}P} &
  \multicolumn{1}{l|}{\cellcolor[HTML]{00FF00}P} &
  \multicolumn{1}{l|}{\cellcolor[HTML]{FFFF00}A} &
  \multicolumn{1}{l|}{\cellcolor[HTML]{00FF00}P} &
  \multicolumn{1}{l|}{\cellcolor[HTML]{00FF00}P} &
  \cellcolor[HTML]{00FF00}P &
  \multicolumn{1}{l|}{\cellcolor[HTML]{00FF00}P} &
  \multicolumn{1}{l|}{\cellcolor[HTML]{00FF00}P} &
  \multicolumn{1}{l|}{\cellcolor[HTML]{00FF00}P} &
  \multicolumn{1}{l|}{\cellcolor[HTML]{FFFF00}A} &
  \multicolumn{1}{l|}{\cellcolor[HTML]{FFFF00}A} &
  \multicolumn{1}{l|}{\cellcolor[HTML]{00FF00}P} &
  \cellcolor[HTML]{00FF00}P \\ \midrule
\textbf{Kavi} &
  3987 &
  \multicolumn{1}{l|}{\cellcolor[HTML]{00FF00}P} &
  \multicolumn{1}{l|}{\cellcolor[HTML]{00FF00}P} &
  \multicolumn{1}{l|}{\cellcolor[HTML]{FFFF00}A} &
  \multicolumn{1}{l|}{\cellcolor[HTML]{FFFF00}A} &
  \multicolumn{1}{l|}{\cellcolor[HTML]{00FF00}P} &
  \multicolumn{1}{l|}{\cellcolor[HTML]{00FF00}P} &
  \cellcolor[HTML]{00FF00}P &
  \multicolumn{1}{l|}{\cellcolor[HTML]{FFFF00}A} &
  \multicolumn{1}{l|}{\cellcolor[HTML]{00FF00}P} &
  \multicolumn{1}{l|}{\cellcolor[HTML]{00FF00}P} &
  \multicolumn{1}{l|}{\cellcolor[HTML]{00FF00}P} &
  \multicolumn{1}{l|}{\cellcolor[HTML]{00FF00}P} &
  \multicolumn{1}{l|}{\cellcolor[HTML]{00FF00}P} &
  \cellcolor[HTML]{00FF00}P \\ \midrule
\textbf{Pri} &
  1760 &
  \multicolumn{1}{l|}{\cellcolor[HTML]{00FF00}P} &
  \multicolumn{1}{l|}{\cellcolor[HTML]{00FF00}P} &
  \multicolumn{1}{l|}{\cellcolor[HTML]{00FF00}P} &
  \multicolumn{1}{l|}{\cellcolor[HTML]{00FF00}P} &
  \multicolumn{1}{l|}{\cellcolor[HTML]{FFFF00}A} &
  \multicolumn{1}{l|}{\cellcolor[HTML]{00FF00}P} &
  \cellcolor[HTML]{00FF00}P &
  \multicolumn{1}{l|}{\cellcolor[HTML]{00FF00}P} &
  \multicolumn{1}{l|}{\cellcolor[HTML]{FFFF00}A} &
  \multicolumn{1}{l|}{\cellcolor[HTML]{00FF00}P} &
  \multicolumn{1}{l|}{\cellcolor[HTML]{00FF00}P} &
  \multicolumn{1}{l|}{\cellcolor[HTML]{00FF00}P} &
  \multicolumn{1}{l|}{\cellcolor[HTML]{00FF00}P} &
  \cellcolor[HTML]{00FF00}P \\ \midrule
\textbf{Dev} &
  3885 &
  \multicolumn{1}{l|}{\cellcolor[HTML]{00FF00}P} &
  \multicolumn{1}{l|}{\cellcolor[HTML]{FFFF00}A} &
  \multicolumn{1}{l|}{\cellcolor[HTML]{00FF00}P} &
  \multicolumn{1}{l|}{\cellcolor[HTML]{00FF00}P} &
  \multicolumn{1}{l|}{\cellcolor[HTML]{00FF00}P} &
  \multicolumn{1}{l|}{\cellcolor[HTML]{00FF00}P} &
  \cellcolor[HTML]{FFFF00}A &
  \multicolumn{1}{l|}{\cellcolor[HTML]{00FF00}P} &
  \multicolumn{1}{l|}{\cellcolor[HTML]{00FF00}P} &
  \multicolumn{1}{l|}{\cellcolor[HTML]{FFFF00}A} &
  \multicolumn{1}{l|}{\cellcolor[HTML]{00FF00}P} &
  \multicolumn{1}{l|}{\cellcolor[HTML]{00FF00}P} &
  \multicolumn{1}{l|}{\cellcolor[HTML]{FFFF00}A} &
  \cellcolor[HTML]{00FF00}P \\ \bottomrule
\end{tabular}
\caption{An illustration of the roster plan. A refers to ‘absent’ indicating off-days and ‘P’ refers to ‘present’ indicating work-days. Every annotator received 6 days off each month. If they got 1 day off in the first week, they would get 2 days in the second week and 1 day again in the third week and so on. However, there was no guarantee that these days would be consecutive or fall on weekends.}
\label{tab:my-table2}
\end{table}

It wasn’t just people that were counted. Tasks too were counted. Counts had to be made to take stock of each day’s work: how many tasks were received from the client, how many were completed, how many were remaining to be carried over to the next day. Such details were compiled into end-of-day reports (EoD) for the client. These reports served to convey the progress made that day and account for the hours passed. 

In each project, managers and TLs were entrusted with this responsibility to compile daily counts into EoD reports\footnote{Besides the EoD reports, there were also weekly, monthly and quarterly reports that aggregated and analysed a wider range of counts for the weekly, monthly and quarterly business review (respectively referred to as WBR, MBR and QBR) meetings with the client.}. This included tabulating the number of tasks newly received on the day, carried over from the previous day, completed and pending. This was supplemented with the annotator count for the day and how many hours they spent on the tasks. They retrieved and tallied up these numbers from the annotation tools used in that project, client emails and spreadsheets used for project management. Taken together, the task counts and annotator counts accounted for the day’s progress or the lack of thereof. For instance, on one occasion, the annotation tool was not functional for 3 hours, affecting the day’s progress. This was duly noted in the EoD report, as a one-line note below the table of counts. Through such prompt reporting and accounting for time, annotation centres sought to inspire the client’s confidence in their professional conduct of annotation work. 

Both roster planning and EoD reporting involved enumerating people and the work they did. In the case of roster planning, counting was essential to organise and manage day-to-day annotation work. With EoD reports, counting’s necessity lies not in being indispensable to conducting annotation work but in being a compulsion of conducting the work within the transnational outsourcing supply chains. The necessity to count and account made it an unquestioned routine, a taken for granted practice that was a part of all annotation projects.  In this way, counting tasks and people were intricately woven into everyday annotation work practices. 

\subsection{Counting as the de-facto language}

Woven into everyday annotation work, counting forms a vocabulary to describe annotation, be it in roster plans or EoD reports. Using this vocabulary sets further counting in motion, shaping how annotation is thought of, understood and conceptualised. To illustrate this, in this section, we describe three numbers that were important in the counting vocabulary: hourly targets, average handling time (AHT) and goodwill scores. By tracing the production and use of these specific counts, our aim is to bring attention to the dominance of counting as a logic, and its recursive role in shaping annotation work practices. 
 
The EoD reports described in the previous section only provided the overall number of tasks received, completed and pending; it did not include any further breakdown of work done by the hour or each annotator in the team. Yet, in both data centres we visited, those details were collected and closely monitored to ensure smooth progress could be reported in the EoD figures. Almost every project used an hourly tracker, often maintained as a spreadsheet. Here, each annotator logged the number of tasks they finished each hour. TLs typically marked the passing of each hour with announcements like, “\emph{Update your count in the [Google] Sheet. Every hour you need to complete at least 25 events. Make sure your hourly count is updated properly; only then the work is complete. And only if you have done 25 per hour, you can leave [for lunch].}” This was a cue for annotators to both log their past hour’s count as well as ensure they achieved their hourly target. And TLs tracked this closely as a way to pre-empt problems and minimise disruption in annotation work. For instance, with hourly counts, they identified technical issues or glitches early on— if the annotation tool was slow or an annotator had doubts, the resulting low hourly count would bring it to their attention without delay. There were also other kinds of meanings that TLs associated with low hourly counts: an annotator’s low hourly count at the start of the day could mean rushed, error-prone work later. 

\begin{displayquote}
“With the hourly tracker, I can see how each member is performing. For example, if they matched only 30-40 products in the morning and the target for the day is 250… How will they finish the remaining in the afternoon? Errors only will be more then. At that time, I can see here and identify why more errors are coming”

\rightline{- Das, junior TL in the product matching project, Veera DF} 
\end{displayquote}

But not all annotation projects could be cast in terms of hourly counts and targets. For instance, in the live support for the cashierless checkout project, the flow of tasks was not steady or entirely predictable. It depended on many factors including the crowd at the checkout-free stores, device “health”\footnote{ Device health referred to the status of cameras and weight sensors used by the AI system to detect  when a product was picked for purchase. If the sensors misfired or cameras failed, the AI system would not be able to accurately detect shopper purchases.} of sensors and cameras, lighting in the stores, and the ability of the AI system to capture and convert the shoppers’ activities into a shopping cart. This meant that an annotator could receive 6 live support tasks in the first hour, 20 in the second hour and 15 in the third. Here, the hourly count neither offered a granular view of task completion rates nor did it lend well as a target. The alternative count used in this project was average handling time (AHT). 

At the end of each day, an internal support team at the Chandra DF annotation centre downloaded all the live support tasks performed that day, aggregated the time taken per task and computed the AHT by task type and per person. This was shared as a report the next day with the TLs of all shifts. The report tabulated the overall AHT by task type and compared them against the baseline AHT figure agreed upon with the client. As illustrated in Table 3, each AHT value is highlighted in green or red depending on whether the baseline AHT was “met”. It also includes the AHT trends from the past week to analyse continuities and discrepancies. A similar report is created tabulating the AHTs of each annotator and colouring them green or red by comparing their AHT against baseline values.

\begin{table}
\begin{tabular}{@{}|l|l|l|l|l|l|l|l|l|@{}}
\toprule
\textbf{Task Type} &
  \textbf{\begin{tabular}[c]{@{}l@{}}AHT\\ Baseline\end{tabular}} &
  \textbf{\begin{tabular}[c]{@{}l@{}}July \\ 13\end{tabular}} &
  \textbf{\begin{tabular}[c]{@{}l@{}}July \\ 12\end{tabular}} &
  \textbf{\begin{tabular}[c]{@{}l@{}}July \\ 11\end{tabular}} &
  \textbf{\begin{tabular}[c]{@{}l@{}}July \\ 10\end{tabular}} &
  \textbf{\begin{tabular}[c]{@{}l@{}}July \\ 09\end{tabular}} &
  \textbf{\begin{tabular}[c]{@{}l@{}}July \\ 08\end{tabular}} &
  \textbf{\begin{tabular}[c]{@{}l@{}}July \\ 07\end{tabular}} \\ \midrule
Interaction labelling &
  1.1 mins &
  {\color[HTML]{6AA84F} 0.8 min} &
  {\color[HTML]{6AA84F} 0.9 min} &
  {\color[HTML]{6AA84F} 0.8 min} &
  {\color[HTML]{6AA84F} 1 min} &
  {\color[HTML]{6AA84F} 0.8 min} &
  {\color[HTML]{6AA84F} 0.9 min} &
  {\color[HTML]{6AA84F} 1.1 min} \\ \midrule
Person matching &
  1.2 mins &
  {\color[HTML]{6AA84F} 1.1 min} &
  {\color[HTML]{6AA84F} 1 min} &
  {\color[HTML]{6AA84F} 1.2 min} &
  {\color[HTML]{6AA84F} 1.1 min} &
  {\color[HTML]{6AA84F} 1 min} &
  {\color[HTML]{FF0000} 1.3 min} &
  {\color[HTML]{6AA84F} 1.2 min} \\ \midrule
Check billing &
  4.9 mins &
  {\color[HTML]{FF0000} 7.2 min} &
  {\color[HTML]{FF0000} 5.6 min} &
  {\color[HTML]{FF0000} 6.8 min} &
  {\color[HTML]{FF0000} 5.7 min} &
  {\color[HTML]{FF0000} 6.8 min} &
  {\color[HTML]{FF0000} 6.6 min} &
  {\color[HTML]{FF0000} 5.8 min} \\ \midrule
Confirm shopper &
  0.9 mins &
  {\color[HTML]{FF0000} 1.1 min} &
  {\color[HTML]{6AA84F} 0.7 min} &
  {\color[HTML]{6AA84F} 0.5 min} &
  {\color[HTML]{6AA84F} 0.8 min} &
  {\color[HTML]{6AA84F} 0.6 min} &
  {\color[HTML]{6AA84F} 0.5 min} &
  {\color[HTML]{6AA84F} 0.5 min} \\ \bottomrule
\end{tabular}
\caption{An illustration of the overall AHT report for a week in July 2022. The task type details and the AHT values have been altered to maintain project confidentiality. The AHT values in green indicate that they were lower or equal to baseline values, while those in red indicate AHTs that exceeded baseline values.  }
\label{tab:my-table}
\end{table}

Time taken to complete each task was not the only detail to be tracked. The mistakes that annotators committed in tasks were also similarly logged, monitored, compiled and counted. Every time a task’s verdict was changed during quality check, the original verdict was marked as an error, for the original annotator. Each error was thus mapped on to an annotator and for each annotator, there existed a compilation of errors they had committed, later translated into accuracy rates. 

Hourly counts, AHTs, error trackers, accuracy rates, attendance trackers, roster plans - all of them together were combined and weighed to evaluate annotator performance, on a scale of 1 - 4. Everything that could be counted was factored into this score, a master count. And even aspects that didn’t lend well to counting were made countable. The performance evaluation score had three parts to it - productivity, accuracy and “goodwill”. The latter was calculated with questions such as How compliant was the annotator? Were they proactive in taking on tasks? Did they support their teammates? Did they frequently request time off or take unplanned leaves? Were they “flexible” in adapting to the demands of annotation? The answers to these questions could be direct, elaborate, complicated or conflicting but they eventually boiled down to one number - the goodwill score. 
 
Annotation work, then, involved more than just being productive and accurate. It needed compliance, creativity, enthusiasm, flexibility and adaptability from annotators. Along with productivity and accuracy, all of these factors differentiated each annotator’s performance. But to be accounted for within a practice (ie., annotation) that had developed a fluency for counting, all of these aspects also needed to find words (or numbers actually) within that vocabulary. The goodwill score was that catch-all count to account for everything that couldn’t be captured by accuracy and productivity. 

The range of trackers, targets and scores elaborated in this section reveal the dominance of  counting and its logics in organising and managing annotation. They draw on and extend the counting vocabulary seen in periodic reporting and accounting practices (described in the previous section), making counting the common, de-facto language to describe, discuss and make decisions in annotation. The day-to-day activities in annotation were seldom conducted or described without using the vocabulary of counting. However, as seen with the goodwill score, this required even those aspects that didn’t lend well to enumeration to be counted in order to be accounted for.

\subsection{Counting as negotiating}
In day-to-day data annotation work, counting was necessary, pervasive and dominant. Yet, even though counting was taken for granted, the numbers it produced (be it shift headcounts, AHTs or hourly targets) were not a given. They were a result of deliberate negotiations. How long does a task take to complete? What was a reasonable hourly target? How many annotators were needed in a shift? How long would a team of 10 require to annotate a 50K dataset? These questions were pertinent to annotation practice and yielded specific counts such as the baseline AHT, hourly targets and shift headcounts. The counts were produced through benchmarking, a process of deliberate negotiation between annotation teams and their requesters to define pace, productivity and time commitments for annotators to comply with, in their everyday work. In this section, we elaborate on the benchmarking process, drawing attention to the negotiations and contestations, shaped by power asymmetries and opposing goals, and their implications for other forms of counting. 

Benchmarking was a periodic exercise carried out for each new type of task in a project, to compute and agree upon the AHT or hourly target. It involved the annotation team working on a dataset to assess and compute how many tasks they could each complete in an hour’s time. This helped evaluate the complexity of the task, and inform the target numbers for subsequent datasets of a similar kind. 

\begin{displayquote}
“What we do is, we see how much each person is able to do in an hour, right? If we take the average of that, we will be able to communicate that to the client as the benchmarking result - that we are going to set this as the target for this set [i.e., file]. Once they agree, we’ll be able to start production.”

\rightline{- Shankar, assistant manager, checkout-free shopping project } 
\end{displayquote}

“Once they agree” was the key. The benchmarking exercise was carried out for the two parties to come to an agreement about how much annotation work could be completed within a day or an hour; how much time each task needed. Annotators submitted this information to their client for approval. It was only (if and) after the client approved the average handling time or hourly target, could the annotators begin processing the dataset for production. Typically, the annotators benchmarking the dataset were the first ones to lay eyes on the dataset and gauge the nature and complexity of the tasks in it. Yet, it was not uncommon for clients to insist on improving the targets, pushing them higher. 

\begin{displayquote}
“The client will say, ‘this is a new set, the tasks will be a little different in this, so you can do benchmarking first. Last few sets you have done 100 tasks per hour, and based on the annotations, we have trained the algorithm. Now you may be able to do more because the task has changed a bit.’ So, we work on benchmarking and get back on the targets, which they have to accept. If they don’t accept, we will have a call and try to come to an agreement.”

\rightline{- Shankar, assistant manager, checkout-free shopping project } 
\end{displayquote}

In other cases, the insistence could be a lot more direct and harder to negotiate. In a new pilot project at Veera DF, senior TL Raghu, who liaised with the client, recounted the pressures of negotiating targets. Through the benchmarking process, they identified 50 tasks an hour as the realistic target; but the client’s expectation remained at 60-70 per hour. 

\begin{displayquote}
“In the mapping project, we check [through benchmarking] how much [work/tasks] agents in our team can do and give them a count, say 50 per hour. They [the client] may push[back] saying no, you can do 60-70 per hour. What they say is… we also work out some samples while giving the project. Those are the pressures while handling clients.”

\rightline{- Raghu, senior TL, Veera DF } 
\end{displayquote}

While on one level, annotation teams were pushing back unrealistic targets, their senior managers would simultaneously be looking for and creating opportunities to expand the project. On monthly and quarterly review calls with the clients, they would pitch for taking on more work – this could be increasing headcount, forming new teams or getting allied projects. Annotation projects rarely started with a large team or a big headcount. They almost always started small, and grew over time as they consistently delivered high quality datasets, earning the trust and appreciation of the client. The Veera DF centre began its first ecommerce annotation project for a large corporation in 2012 with a small team of 5. Ten years later, at the time of our study, they continued working with the client on multiple annotation projects employing over 1500 annotators. They obtained the latest project with this client when a PoC (point-of-contact), impressed with their track record, moved teams and was happy to expand their collaboration with a new project. Similarly, in the live support project for checkout-less shopping, the client’s headcount requirements did not include quality control analysts. But the Chandhra DF centre voluntarily conducted quality assurance checks for every task (allocating time and resources for it at its own expense) to help with future negotiations to increase the project headcount . 

\begin{displayquote}
“Internally, we are allocating people and assigning resources to ensure that we are going with good quality. We check the submitted tasks and stand at 99.5\%. That is our audit process. Right now, the client is not that focussed on quality. In the last QBR, we shared our audit process results and gave a proposal — if you assign some QC resources, we will do it dedicatedly and produce a report too. We have given that input, they have to come back. ”

\rightline{- Balaji, manager at Chandra DF  } 
\end{displayquote}

Thus we see that calculations like the time required to complete one task or headcount required in a project were central to organising annotation through a variety of other counts. Annotation teams had to show promise, showcase results, and bid time while making these calculations. The clients wielded significant influence over the calculations and through that, dictating the pace, productivity and scope of annotation work. Benchmarking was therefore not just a process to calculate average handling time or hourly targets but to impose them. Any negotiation in the process was further overshadowed by implications for other kinds of counts and goals. Benchmarking datasets and making sales pitches were two separate activities, with very different purposes, conducted at different levels of organisation, towards opposing goals. With benchmarking, annotators aimed to limit client expectations of productivity; while with sale pitches, they sought to expand the work outsourced to them. These opposing goals were entwined by the numbers they created, championed and contested. Pushing back high hourly counts could turn on efforts to push the project headcount up. Annotation managers negotiating these counts were acutely aware of such knock-on effects. 

\subsection{Counting as falling short}

To organise annotation through counting logics, annotation itself is distilled and reduced to incremental units. These units are designed to afford counting, aggregating and comparing, and to be ready for automated analysis. For instance, to monitor a team, progress would be judged against individual annotators each working on their own tasks, independently. To calculate the time taken per task, granular measurements of time were important. And tasks had to be either right or wrong, accurate or an error, with no scope for uncertainty or ambiguity. Only then could all of the tasks, targets, people, and errors be counted, clocked, aggregated. These numbers then allowed comparisons and rankings to be made so that all could be understood, accounted for and reconciled within a counting logic. However, such a logic of counting does not allow for recognising certain crucial realities of annotation. In this section, we illustrate how counting falls short of capturing the collaboration, pressures and ambiguity in annotation work. 

Counting fell short in accounting for pressures in annotation work and its toll on annotators. Benchmarking exercises commissioned to determine target counts and AHTs did so by computing the time taken to complete new tasks over an hour or so. But this failed to account for annotators’ experiences of working on the same tasks repeatedly over extended periods of time. The AHTs of a new dataset calculated in the first hour of working on it could not capture the fatigue of doing them all day long, potentially for several weeks before the next new type of task was introduced. Annotation teams were well aware of and regularly highlighted the experiences and limitations of repetitive work while setting targets. A target-led approach to annotation necessitated an exercise like benchmarking to gain granular insight into the time taken to complete one task. By this logic, time involved in annotating could be best understood by zooming in, not out. The more granular the time unit, the better or more precise. This in turn meant that sampling any one hour of annotation would be representative of every hour of annotation to follow. If 50 tasks were done in the first hour, this logic implied that 50 could be (and therefore should be) done each hour. There were no numbers to render each hour’s toll on the next.

\begin{displayquote}
“What does not get acknowledged [by clients] is that our annotators have to do this task everyday… Yes, in one hour you can do these many tasks but when you do that same thing repeatedly for so long, you cannot keep delivering the same count. It is not possible. ”

\rightline{- Kantha, senior management at the headquarters } 
\end{displayquote}

Similarly, collaboration in annotation could not be counted; it did not align with a logic where each task counted towards one person’s hourly count, target, accuracy rate and errors. A common practice among annotators was to tilt their monitors towards their peers sitting next to them and ask what they thought of a particular task: is this shopper picking an apple or putting it back? Is this a bottle of Coke or Dr. Pepper? Are these two products a match if the colour names are different? Annotation verdicts often had to be provided over ambiguous contexts, with limited information. As they worked with blurry video footage, culturally unfamiliar products and complex scenarios, annotators constantly sought the wisdom of their colleagues to arrive at a verdict. They helped one another, working together to solve each other’s problems; a goodwill integral to accurate annotations. 

The most striking instance of such collaborative annotation was observed in the till-less shopping project, when a group of 3-4 annotators all worked on one task - verifying the shopping cart of an adult who entered the store with three children. The children raced each other between aisles, picking random products that caught their attention. The adult accompanying them rejected several of their additions to the cart, removing them with disapproval. As this group exited the store a few minutes later, annotators in Chandra DF had to verify if their final bill (auto-generated by the AI system) was accurate, charging them only for the items they left with. For this task, each annotator followed the video of one child going through the store to carefully trace all items they successfully added to the cart. Finally, having traced each member of the group, they collectively decided if the final cart was accurate. 

These collaborative efforts could not be captured by the established modes of counting in Chandra DF. Such counting rested on the premise of a 1:1 mapping between verdicts and annotators—it assumed that every verdict could be associated with one annotator. The elaborate efforts in counting, the purposes they served and the meanings produced through them were only countable if each task was seen as being done by one annotator. It did not leave room to account for collaborations in annotation. 

Besides overlooking such realities of annotation practice, counting also interfered with them.  In pursuit of a clean count with no errors, annotators were motivated to skip tasks which were complex or ambiguous to annotate. For instance, in the product matching project at Veera DF, annotators encountered many challenges in their task of determining whether two products were the same or not. On one occasion, Teju, an annotator, struggled to figure out whether the base product was a movie or a game. She poured over the description, which seemed similar to other video game descriptions she had come across. She was not sure if the movie itself involved a game or perhaps the movie was popular and turned into a game as well. She consulted her colleagues, who took turns to read through and compare the two products. Nobody was entirely sure. That’s when Ram, another annotator, recommended, “\emph{It’s not clear right? Just skip this task. If our accuracy drops, they will call for a meeting and lecture us! Do you want all this? Just skip the task.}” Low accuracy and hourly counts reflected poorly in performance evaluation and could result in the annotator being pulled up for further scrutiny of work. Rather than risk this fate, Teju was advised to resort to the easier alternative - clicking on ‘Skip Task’, to avoid a bad count. 

In the ambiguous, challenging tasks lay opportunities to better understand the world that is being modelled, to build a richer picture and to train a more robust system. Rather than being encouraged to pursue, identify, develop and share insights on annotating ambiguity in the real world, annotators felt dissuaded to deal with anything that jeopardised their count. Thus, within a counting regime that enumerates tasks, errors, accuracy and productivity, opportunities that threaten the count are skipped and lost.

\section{Discussion}
The quantification of work is by no means a new theme in CSCW. There is a rich body of scholarship that closely examines the rising prominence of metrics in the datafied workplace \cite{10.1145/3544548.3580950, 10.1145/3555531, 10.1145/2750858.2807528, 10.1145/3025453.3025510, 10.1145/3544548.3581376, 10.1145/2818052.2855514, 10.1145/3491102.3501866, 10.1145/3432949, 10.1145/3476060, 10.1145/3555177}. Our paper contributes to the existing body of knowledge by delving into the quantification of data annotation work. We specifically focus on the act of counting within annotation tasks, as it plays an integral role in the everyday practices of annotation work. As we witness a surging interest in the adoption of AI, annotation — a foundational aspect of AI production — warrants close scrutiny, particularly the ways in which it is structured and conducted. Our work offers an up close view of how data annotation is constituted through counting practices and reflects on its implications for AI. 

Through our empirical study of data annotation, we unpack the central role of counting. The different counts, along with their underlying logics and practices described in the sections above, indicate the extent to which counting is woven into annotation. The day-to-day activities in annotation were seldom carried out or described without using the language of counting. This pervasiveness of the count in annotation is then an ordinary feature of the data work, so much so that the practices and structures involved in producing numbers have become taken for granted and indeed normalised. It becomes hard to imagine the work of annotation any other way. Yet, our findings detail counting practices that are far from passive or inert—mere representations of labour or tools to enumerate productivity. Rather, counting exerts a definite impact on the ways annotation work is conducted, organised, managed, and valued. The assumed countability of virtually everything associated with annotation shapes the work and ultimately impacts what is delivered to those developing AI models and systems.

Building on this empirical evidence and putting it in conversation with the sociological and socio-technical scholarship on counting \cite{porter1996trust,verran2001science, rose1991governing, bowker2001pure, starr1987sociology, orlikowski2002s, 10.1145/1958824.1958861}, in this section, we examine the totalising counting practices in annotation as part of a \emph{regime of counting} \cite{bowker2001pure}. In such a regime, we call attention to the structuring \cite{orlikowski2002s,10.1145/1958824.1958861} work in operation through numbers and the resulting tensions and erasures in counting. We begin by examining counting as a structural activity that situates control and authority amongst a few actors. Clients, as we’ve noted above, wield a disproportionate influence over the processes and outcomes of annotation through counting. Consequently, they control what gets counted and accounted for. The fallout of their priorities and interests can then be traced through the tensions and erasures in counting. Through these themes, the analytical frame of regimes of counting provides an analytical orientation to examine who exerts control over annotation through quantification and to what ends they do so. In closing, we reflect on the implications of this structural arrangement for annotation as well as for the broader development of AI systems.

\subsection{The structuring work done by numbers}

As the de facto language and dominant logic of annotation, counting produced many numbers, at different granularities, for different purposes. Each annotator’s hours spent at work, minutes spent on a task, errors per dataset, or tasks per hour were enumerated to motivate productivity, maintain accuracy, monitor and evaluate performance, negotiate workloads, expand business, and comply with deadlines. Underlying the many meanings that the various counts took on is the normative assumption that counting is appropriate and adequate to make sense of annotation. This view relies on and feeds into the notion that counting is a methodical, rigorous practice almost an inevitable solution for organising work and human labour. But this overlooks the structuring work done by numbers and counting, that numbers and their attendant counts are \emph{not} at all neutral.

We see this, for instance, in roster planning, where work is divided per person over time, neatly compartmentalised as shifts. Here, work is configured using numbers. This also necessitates further counting—taking attendance, planning the roster, accommodating leaves and so on. To meet each shift’s expected headcount, while also weaving in off days, required reconciling the two counts. One number had to be accommodated without upsetting the other. Once the roster was drawn up, annotators were free to swap their off-days with colleagues, so long as it did not disturb the required count for the shift. In this subordination of one count to another, we begin to see counting as a structural activity, that is an activity that organises and orders activity and in doing so ascribes normative values and judgments.

Annotation is replete with such reconciliations and tensions within its many counts. We see it more explicitly in benchmarking where counting meant to determine effort and time required to annotate datasets is a contested process. Though the annotators were the first ones to review the complexity of a dataset, it was the clients who held the final say on the time needed to annotate it. Far from being an objective exercise to measure average handling time, benchmarking was an exercise of power and unequal negotiation. Targets born out of such benchmarking negotiations allowed the clients to exert control over the annotation process. This control was further reasserted at the managerial level through hourly counts, that enforced targets on annotators each day, and the ‘goodwill’ score, that quantified and standardised the (intangible) ways in which annotators conducted themselves and conducted their work. 

In this way, their contribution could be counted, and thereby accounted for in a manner that serves to assert conformity, and locate control amongst a few actors. Within the transnational supply chain of AI and data work, counting regimes aid, as they historically have \cite{godfrey1996accountability, oldroyd1997accounting, mcdonald2005using,mennicken2019s, cohn1984census,cohn2020colonialism, kalpagam2000colonial, samarendra2011census, van2004establishing}, the assertion of authority over processes distant from the centres of power. They throw a veil of neutrality and objectivity over processes steeped in power asymmetries. 

\subsection{Tensions and erasures in counting}
Counting regimes do not merely ‘represent’ an existing reality, they constitute it \cite{rose1991governing,latour1987science,hacking1983representing,nelson2015counts}. Counting creates a “clearing” \cite{power2004counting} within which annotation can be understood, imagined and acted upon. EoD reports, benchmarking, performance metrics; each of these produces such clearings to make sense of, manage and inform decision-making in annotation. What gets counted (and accounted for) in these clearings then constitutes annotation in specific ways. As we have seen, benchmarking does not account for the effects of doing the same type of task repeatedly over extended periods of time. The clearing forecloses the possibility to see the toll each hour’s work takes on the next. Similarly, the efforts to meticulously record annotation work narrowly through individual annotator’s counts and accuracy rates overlook the collaborative nature of annotation work. The counting regime operates and is enacted through a logic prioritising individual entities over collective and emergent relations. The creation of a counting clearing, then, sets a naturalised position from which to view annotation practices, which itself makes counting harder to contest \cite{bowker2000sorting}. Any challenge to the count must be defined in the same vocabulary, further reinforcing its normative power \cite{power2004counting}. 

Annotation work, in its broadest terms, is a processing and distilling of complex social realities; it involves the translation of subjective human discretion into unitary tasks that can only be known and rendered as numbers. As Starr \cite{starr1987sociology} has argued, this reduction in complexity through counting is not ideologically innocent. Rather, it is an intentional choice that is indicative of the interests and priorities of the clients who dictate what is counted, by what methods, and to what ends \cite{rose1991governing}.

Seeing through the prism of this regime of counting, raises significant implications for the workers engaged in data annotation and, in turn, for the AI supply chain. What we argue, in closing, is that the structuring work done through enumeration, the attendant tensions and erasures and how both are placed within a wider frame of a regime of counting have implications for:
\begin{enumerate}
\item The value and valuing of annotation work
\item The recognition of human discretion necessitated in annotation work, and
\item Efforts to introduce accountable and more just AI systems.
\end{enumerate}

In each of these cases, the regime of counting is found to exert control, authority and power in annotation and data practices. Although seeming to smooth over the practicalities of annotation work, this counting regime amplifies the tensions and ultimately further problematizes issues that are already seen to be significant obstacles for a tech sector that is seeking to act more responsibly \cite{10.1145/3449081, 10.1145/3298735, 10.1145/3334480.3375158, 10.1145/3544549.3583178, 10.1145/3491101.3516502}. 

\emph{The value and valuing of annotation work}—The regime of counting enacted through a variety of managerial, procedural and task-based processes (only some of which we have detailed above) makes annotation legible in a distinctive way. For instance, in the periodic reports to their clients, annotation teams shared regular updates about the status of their work. This included reporting progress in terms of tasks completed and pending, trends in AHT over a month or quarter to justify slumps and highlight improvements, the time spent on different types of tasks with footnotes on efforts to optimise them, and trends in accuracy rates followed by plans to improve them. Through these different counts, the reports, for the most part, record project status in terms of annotators’ time and efficiency at work; the actual contributions and insights of annotators do not find space here. The definition, scope and methods of an enumeration, all embody the expectations, beliefs and concerns of those dictating the work \cite{rose1991governing}. In the transnational supply chain of data work and AI, the periodic reports record annotation work in specific terms that reflect the interests of the outsourcing requesters and their impressions of annotation work.     

We see then that procedures are enforced that enumerate annotation tasks in terms of time or discrete units. Thus they are cast as a series of identical, repetitively accomplished tasks. This bears significant implications for how this labour is valued. Establishing and formalising this work as repetitive affects how annotators are compensated in their jobs as well as their career trajectories and future prospects. Earlier work pointing to the low-status accorded to data work \cite{wang2020please, chandhiramowuli2023match, sambasivan2021everyone} and its unfair wage structures \cite{10.1145/3555561, posada2022embedded, ismail2018engaging} show that despite being indispensable to AI production, data annotation is considered to be simple, standardised and low-skilled. The counting regime is then implicated in feeding on and feeding into this characterisation \cite{10.7551/mitpress/6352.003.0013} of data annotation, deepening the fault lines of disparity between model work and data work \cite{wang2022whose, martin2014being, sambasivan2021everyone}. This reinforces the importance of calls to promote the value of annotation and recognize it as integral to the wider AI supply chain \cite{10.1145/3555561, sambasivan2021everyone}. In particular, it highlights the importance of surfacing the limits of counting in the tools and structures that pervade data work.

\emph{The recognition of human discretion necessitated in annotation work}—The regime of counting in annotation also has implications for the AI systems it feeds into. Casting annotation as a simplistic and inert compilation of tasks contributes to a wider failure to appreciate the larger role and contribution of annotation to AI \cite{gray2019ghost,crawford2021atlas}. Specifically, the organisation and design of tasks and their enumerated measures of productivity and quality, diminish the importance of human judgement or discretion needed and indeed sought after in many annotation tasks. Take, for example, the task of product matching described above (in section 4.4.) where the role of the annotator is to resolve the ambiguity in aspects like colour names or product types to identify the right match for products. AI systems failed to do this with reasonable confidence, necessitating annotator involvement. Their input was essential to make product matching robust so that it can inform ecommerce pricing decisions. Yet, we see annotators hesitant to engage with ambiguous tasks, where their input is most valuable, weighing its implications for their error counts and performance scores. Ambiguity and complexity in annotation tasks present opportunities to enrich the AI systems they feed into but this is set against a risk of jeopardising performance and error counts.

We thus see how, within a regime of counting, a presumed annotation quality is enforced by penalising variance and rewarding high “accuracy” and productivity among annotators. However, the operational and managerial emphasis placed on meeting enumerated targets as opposed to capitalising on human judgement and discretion can actually run counter to what AI systems are seeking to achieve. By subordinating quality under measures of productivity, we risk losing the possibility to engage holistically with ambiguous tasks and capture the complexities in datasets (and the real world they aim to represent). The implications for AI systems here would appear to be datasets developed that prioritise high throughput and target volume counts rather than annotators’ opinions or judgements. Perversely, the end result risks data-driven AI models derived from people trying to behave in more machine-like rather than human ways. Posada \cite{Posada2023} argues that improving data quality cannot be dissociated from improving working conditions and labour rights. Our study bolsters this call and specifically identifies the need to dismantle reductive and punitive counting regimes in annotation as an important part of improving labour conditions and data quality in annotation.   

\emph{Efforts to introduce accountable and more just AI systems}-In addition to the implications discussed so far for the annotation workers and AI systems, counting regimes also have broader implications for efforts directed towards improving fairness, bias, diversity and accountability in AI systems. These implications do not arise directly from specific counts or counting practices described in the findings. Rather, they become evident by taking a more holistic view of annotation as organised within counting regimes. As elaborated above, counting logics feed into an imaginary of annotation work as a set of reductive, standardised, homogenous tasks, and annotators as distant\footnote{By distant, we mean both geographically far away from centres of AI model development as well as structurally removed from the processes of AI system building.} actors carrying out repetitive work. This imaginary overlooks the unique position annotators occupy—by virtue of closely interacting and incrementally adding to the datasets, they are well placed to develop a deep understanding of the datasets and the context for which they annotate. Across the projects we observed, annotators became highly acquainted with the distant roads, cities, and stores, unfamiliar products they annotated. These annotators often spent months, if not years, working on the same project, making them the most familiar with the datasets of a specific AI system. Given the right conditions, they could potentially play a pivotal role in scrutinising dataset aspects like coverage, diversity, and representation, contributing to fairness, bias mitigation, and accountability efforts. 

However, the prevailing counting logics entrenched in data work significantly hinders the realisation of this possibility. Within the regimes of counting, annotators are often deprived of the autonomy to explore datasets holistically or offer their insights. Rather than seeing them as experts on the data, the close scrutiny on accuracy rates treats them merely as potential sources of errors that need monitoring. Rather than inviting their observations and insights on the data they dedicate extensive hours to annotate, the obligation to submit periodic reports reduces their role to tallying task counts. And rather than fostering opportunities to take a holistic view of datasets, granular calculations of task duration becomes an elaborate ritual they must engage in. These counting regimes not only constrain the imaginative potential of annotation but also squander valuable opportunities to enhance the richness of datasets and AI systems. Moreover, the dominance of counting as an organising principle extends beyond data annotation, permeating efforts aimed at addressing critical issues such as accountability, fairness and diversity. Recent studies \cite{kapania2023hunt, costanza2022audits,gadiraju2023wouldn,barrett2023skin}  show that questions of accountability, fairness, harm mitigation, representation and diversity tend to be rendered and addressed through the logics of counting, with much emphasis and effort focused  on devising  measurable strategies for ‘de-biasing’ models or achieving greater diversity.

By paying close attention to the role of counting in both the missed opportunities to enhance datasets and the efforts to rectify resulting issues, we discern an intricate connection between them. The corrective endeavours aimed at tackling these larger concerns—such as accountability, bias, fairness, and diversity—often employ the very same counting logic that underpin the problems they seek to resolve. Thus, we see counting regimes and the presumption of \emph{total countability} confronting their own limits. It would seem hard here to see accountability being fully addressed when those measures are inherently rooted in the way an AI model and ultimately, the entire system have been constructed. Success (or failure) can only be assessed and evaluated within the parameters of the system, and thus broader conceptions of fairness, bias or diversity are at best a step removed and at worst obscured by a regime that is predisposed to the count. Addressing the complexities of fairness and bias in AI may necessitate a shift from seeking remedies that operate within the regime of counting and instead confront the structural logics and orderings that sustain the regime.

\section{Conclusion}

In this paper, we examine the work of data annotation. We focus on the role of counting in annotation work, and the ways in which managerial control and accountability are woven into wider structures of counting. Key here is a presumption of total countability in annotation, where everything from tasks, datasets and deliverables to workers, work time, quality and performance are seen to be subject to the logics of counting. Drawing on sociological and socio-technical scholarship, we use Bower and Star’s \cite{bowker2001pure} \emph{‘regime of counting’} to develop critical insights into the relations between counting practices and logics surrounding data annotation. We situate counting regimes within the transnational supply chain of AI and data work, showing how clients largely in the Global North exert control and authority over annotation processes, constituting them as reductive, standardised, and homogenous.

We use this analysis to draw out implications for i) how annotation work and workers get valued, ii) the role of human discretion in annotation, and iii) broader efforts to introduce accountable and more just AI systems. Through these implications, we illustrate the limits of operating within the logic of total countability. Offering a counter position, we argue for a view of counting as partial—located in distinct geographies, shaped by specific interests and accountable in only limited ways. This sets the stage for a fundamentally different orientation to counting and what counts in data annotation. It seeks to broaden and enrich the logics that surround and enact annotation work and understand counting to be more than a question of measurement or quantification (and the attendant forms of managerial control), and instead ask what it might be to count annotation as work that can matter and make a difference in efforts to achieve a responsible AI.

\bibliographystyle{ACM-Reference-Format}
\bibliography{refs}

\end{document}